\documentclass{jltp}

\usepackage{graphicx} 

\title{Observation of non-classical rotational inertia \\in bulk solid $^{4}$He}

\author{Motoshi Kondo, Shunichi Takada, Yoshiyuki Shibayama \\
and Keiya Shirahama}

\address{Department of Physics, Keio University, Yokohama 223-8522, JAPAN\\
 }

\runninghead{Motoshi Kondo \textit{et al}.}{Observation of non-classical rotational inertia in bulk solid $^{4}$He}

\begin{document}

\maketitle

\begin{abstract}
In recent torsional oscillator experiments by Kim and Chan (KC), a decrease of rotational inertia has been observed in solid ${^4}$He in porous materials~\cite{nature,JLTP} and in a bulk annular channel~\cite{science}. This observation strongly suggests the existence of `` non-classical rotational inertia''  (NCRI), i.e. superflow,  in solid ${^4}$He. In order to study such a possible `` supersolid''  phase, we perform torsional oscillator experiments for cylindrical solid $^{4}$He samples. We have observed decreases of rotational inertia below 200 mK for two solid samples (pressures $P = $ 4.1 and 3.0 MPa). The observed NCRI fraction at 70 mK is 0.14 \%, which is about 1/3 of the fraction observed in the annulus by KC. Our observation is the first experimental confirmation of the possible supersolid finding by KC. 
\end{abstract}

\section{INTRODUCTION}
It has been proposed theoretically that Bose-Einstein condensation (BEC) and superfluidity can occur in a quantum solid such as solid $^4$He, because of its large zero-point quantum fluctuation~\cite{Reatto,Chester,andreev,leggett,Saslow}. Among a number of theories, Andreev proposed~\cite{andreev} that zero-point vacancies may exhibit BEC. Leggett suggested~\cite{leggett} that superfluidity, named supersolidity, can appear due to quantum atomic exchanges, and it can be detected as a decrease in rotational moment of inertia, which is called non-classical rotational inertia (NCRI). In spite of many experimental attempts, no distinct evidence of supersolidity was obtained. Recently, Kim and Chan~\cite{nature,JLTP,science} performed torsional oscillator studies for solid $^4$He, and observed a decrease in the moment of rotational inertia at temperatures below 200 mK. Their finding strongly suggests that solid $^4$He undergoes a transition to the supersolid phase, and has revived great interest.
The properties of NCRI observed by KC are peculiar: The NCRI fraction is strongly dependent on oscillating velocity. The onset temperature increases with $^3$He impurity concentration, but has surprisingly weak pressure dependence. Since the discovery there have been studies by other experimental means, such as dc flow~\cite{Beamish}, ultrasound~\cite{Kobayashi} and heat pulse~\cite{Aoki}. However, no positive evidences consistent with the torsional oscillator results have been reported so far.

It is therefore still worth reproducing the results of KC using the same torsional oscillator technique. There is plenty of room for detailed studies in the torsional oscillator experiment, such as frequency dependence, effect of sample geometry and crystal quality. Here we report on the first experimental confirmation of the torsional oscillator results by KC. We have made torsional oscillator studies for bulk solid $^4$He sample in a cylindrical cell, which is expected to produce better crystal quality than the annular geometry employed by KC does.

\section{EXPERIMENTAL}
\begin{figure}[t]
\begin{center}
\includegraphics[width=3.0in,clip]{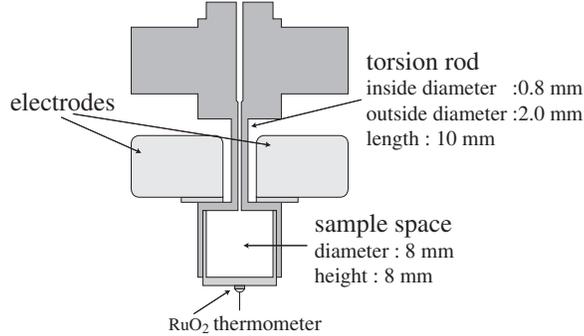}
\caption{The torsional oscillator.}  
\label{oscillator} 
\end{center}
\end{figure}

A schematic view of our torsional oscillator is shown in Fig.~\ref{oscillator}. The hollow torsion rod and the half of the cell were made of a single piece of Be-Cu. The cell was closed by a Be-Cu cap with epoxy (Stycast 2850FT). The sample space has a cylindrical geometry ($\phi\ 8\  \textrm{mm}\times 8 \ \textrm{mm}$). The resonant frequencies $f$ of the empty oscillator and the one when solid $^4$He of 4.1 MPa is formed are 1500.383 and 1496.482 Hz, respectively, at 300 mK. The mechanical quality factor of the oscillator is 2$\times $ 10$^{5}$ at 70 mK. The resonant frequency of our oscillator is roughly $2 \sim 3$ times as high as KC's. We have confirmed that the temperature dependence of $f$ of the empty oscillator does not significantly depend on oscillating velocity in the range that was realized in the experiment. We have acquired the frequency and amplitude data both during cooling runs with cooling speed of about 150 mK/hour, and after fixing temperature for an hour. Both methods gave the same results. 
\par
In order to monitor the pressure of solid $^{4}$He, we have installed a Straty-Adams type capacitance pressure gauge on the massive copper platform, at which the torsional oscillator is mounted. The pressure gauge also contains bulk solid $^4$He and is connected to the oscillator via a 0.1 mm i.d. CuNi capillary. Although this setup does not measure directly the pressure in the oscillating cavity, the measured pressure can be a good indicator of the sample pressure. 

The whole setup was cooled by a dilution refrigerator. Unfortunately the lowest temperature was 70 mK due to some problem in the refrigerator. We have confirmed temperature equivalence between the oscillator cell and the platform by a RuO$_{2}$ thermometer mounted on the cell. 
\par
We have employed commercial $^4$He gas. The concentration of $^3$He is unknown, but is expected to be approximately 1 ppm. Solid $^4$He was prepared by a blocked capillary method. The pressure of the solid sample we report here is 4.10 MPa.  
\section{RESULTS AND DISCUSSION}

\begin{figure}[]
\begin{center}
\includegraphics[width=4.0in,clip]{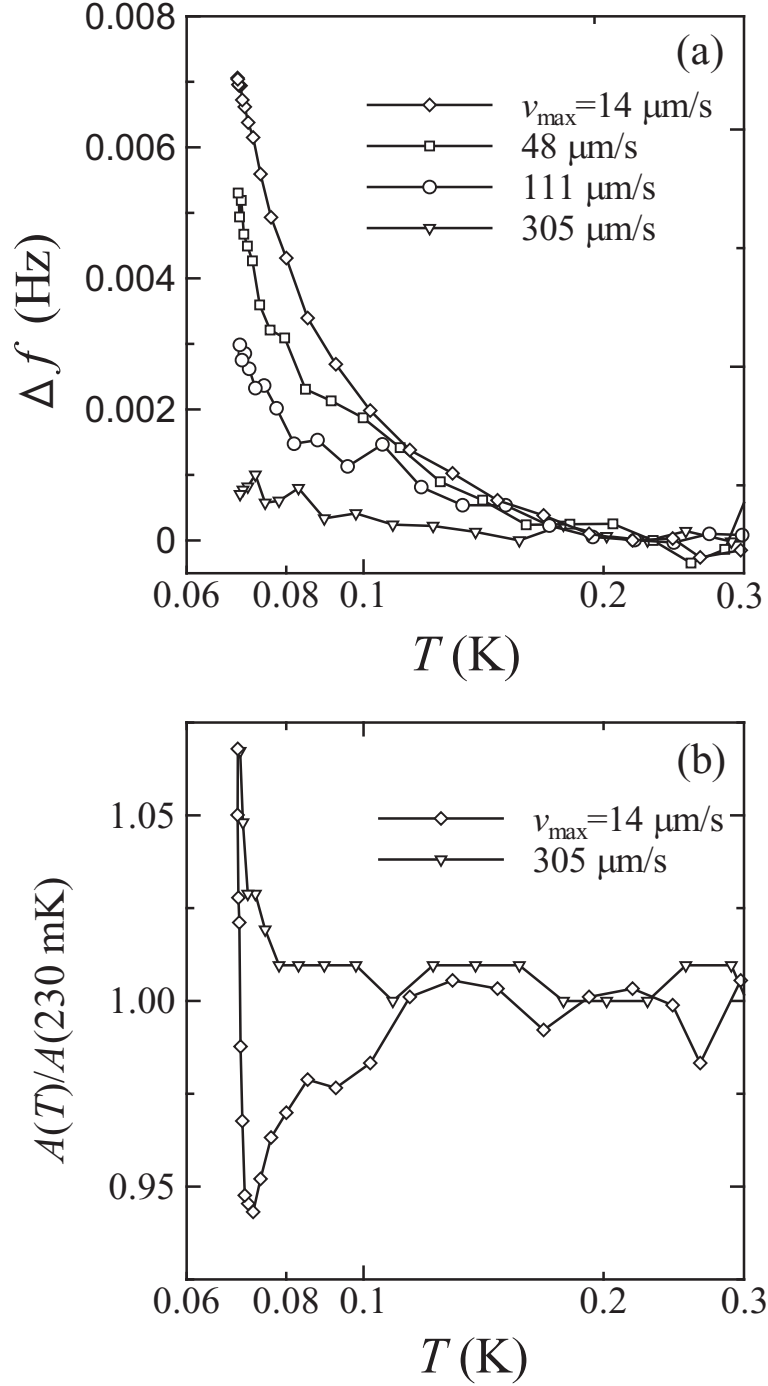}
\end{center}
\caption{(a) The frequency shift $\Delta f$ as a function of temperature at various oscillating velocities $v_{\mathrm {max}}$. After the empty-cell data were deducted, the data were shifted so as to collapse onto a single curve above 230 mK, for easy comparison. (b) Oscillator amplitude normalized at 230 mK, $A(T)/A(230 {\mathrm {mK}})$, for two typical velocities.}  
\label{data} 
\end{figure}

Figure \ref{data} shows typical data sets for a 4.1 MPa solid sample at various oscillation velocities at the rim of the sample, $v_{\mathrm {max}}$. We have observed positive shifts in $f$ below 200 mK. These are clearly shown in Fig. \ref{data}(a), in which the shifts $\Delta f$ were obtained by subtracting the empty cell background from the raw frequency data. $\Delta f$ at $v_{\mathrm {max}} = 14 \mu{\mathrm {m/s}}$ start to increase at about 200 mK, and then increase rapidly below 100 mK. This behavior is quite similar to the observation by KC. In this run we did not observe saturation in $\Delta f$, since the sample could be cooled only down to 70 mK. 
As $v_{\mathrm {max}}$ increases, $\Delta f$ decreases at all temperatures. This velocity-dependent frequency shift is also consistent with the results of KC.

Let us assume that the observed frequency shifts are caused by the mass decoupling of the supersolid component in the solid $^4$He sample. 
The maximum supersolid or NCRI fraction, which is the ratio of the decoupled moment of inertia to the total moment of inertia of the helium sample, is estimated to be about 0.14 \%. This is approximately $1/2 \sim 1/3 $ of the NCRI fraction of bulk solid $^4$He in an annulus at temperature 70 mK~\cite{science}. 

Figure \ref{data}(b) shows the oscillating amplitude normalized at 230 mK, $A(T)/A(230 {\mathrm {mK}})$, at a speed of 14 $\mu $m/s and 305 $\mu $m/s. A sharp dip, i.e. a dissipation peak, is observed in the 14 $\mu $m/s data. The temperature of the amplitude minimum is 74 mK, where the resonant frequency changes most rapidly. As the oscillation speed increases, the dissipation decreases and eventually disappears, as shown in the data at $v_{\mathrm {max}} = 305\ \mu{\mathrm {m/s}}$. The dissipation peak is also one of the key features reported by KC.

These observations are consistent with the KC results, and signify the transition to the possible supersolid phase. We have therefore confirmed at least qualitatively the KC observations.   

\begin{figure}[]
\begin{center}
\includegraphics[width=2.9in,clip]{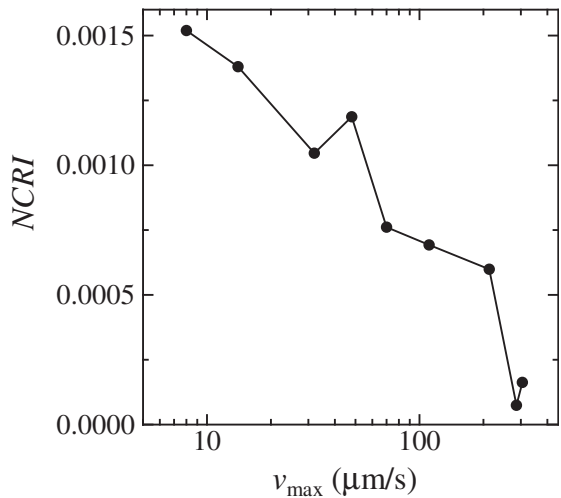}
\end{center}
\caption{The NCRI fraction at 70 mK as a funciton of the maximum rim velocity $v_{\mathrm {max}}$. }  
\label{NCRI} 
\end{figure}

The NCRI fraction strongly depends on $v_{\mathrm {max}}$. We show the dependence on $v_{\mathrm {max}}$ in Fig. \ref{NCRI}.
The NCRI decreases with increasing $v_{\mathrm {max}}$. Although the overall tendency is consistent with the observations of KC, it is difficult to find a critical velocity $v_{\mathrm c}$, which was about $10\  \mu {\mathrm {m/s}}$ in the annular solid experiment reported by KC. In our sample $v_{\mathrm c}$ might be $250 \ \mu {\mathrm {m/s}}$, where the NCRI fraction diminishes rapidly. Further study is needed to elucidate the velocity effect. 

Rittner and Reppy have also confirmed the supersolid behavior in a solid sample at 2.6 MPa with a square-cell torsional oscillator~\cite{reppy}. They have found that the supersolid decoupling and the associated dissipation peak diminish or disappear by annealing of the $^4$He crystal. We have looked for this annealing effect for the present solid sample, by raising the temperature up to 1.8 K, which is slightly below the melting temperature. We could not eliminate the supersolid signal after each of two repeated annealings. The magnitudes of supersolid fraction did not change significantly before and after the annealing. In their experiment, some experimental conditions (pressure, frequency) and sample dimensions differ from ours. In order to clarify the annealing problem, the solid sample has to be prepared more carefully.

We have found that the observed supersolid fraction is smaller than that reported by KC. We attribute this difference to the sample geometry, that is, the cylindrical solid has a smaller supersolid fraction than the annular solid has. This speculation is supported by West, Lin and Chan~\cite{West} who also found smaller NCRI fraction in cylindrical solid samples. This difference may give a clue to elucidate the nature of the observed supersolid phenomena.   

\section{CONCLUSION}
We have observed the frequency shifts and dissipation peaks, which signify possible supersolidity in $^4$He.  
This is the first confirmation of the Kim-Chan experiment. The data are qualitatively consistent with the results reported by KC.
 
\section*{ACKNOWLEDGMENTS}
We are grateful to enlightening discussions with Yuki Aoki, John Beamish, Moses Chan, Eunseong Kim, Harry Kojima, Minoru Kubota, Andrei Penzyev, John Reppy and Masaru Suzuki. 
This work is supported by Grant-in-Aid for Scientific Research on Priority Area "Physics of new quantum phases in superclean materials" (Grant No. 17071010) from The Ministry of Education, Culture, Sports, Science and Technology of Japan.

\end{document}